# A Circulating Biomarker-based Framework for Diagnosis of Hepatocellular Carcinoma in a Clinically Relevant Model of Non-alcoholic Steatohepatitis; An OAD to NASH

Ping Zhou*, Anne Hwang*, Christopher Shi*, Edward Zhu*, Farha Naaz*, Zainab Rasheed, Michelle Liu, Lindsey S. Jung, Jingsong Li, Kai Jiang, Latha Paka, Michael A. Yamin, Itzhak D. Goldberg and Prakash Narayan[1]

\* = equal contribution

1 = corresponding author

Address: Angion Biomedica Corp., 51 Charles Lindbergh Blvd, Uniondale, 11553, New York, USA

 Ph: 516 326 1200

Fax: 516 222 1359

 Email: pnarayan@angion.com

**Abstract**

Although cirrhosis is a key risk factor for the development of hepatocellular carcinoma (HCC), mounting evidence indicates that in a subset of patients presenting with non-alcoholic steatohepatitis (NASH), HCC manifests in the absence of cirrhosis. Given the sheer size of the non-alcoholic fatty liver disease (NAFLD) epidemic, and the dismal prognosis associated with late-stage primary liver cancer, there is an urgent need for HCC surveillance in the NASH patient. In the present study, adult male mice randomized to control diet or a fast food diet (FFD) were followed for up to 14 mo and serum level of a panel of HCC-relevant biomarkers was compared with liver biopsies at 3 and 14 mo. Both NAFLD Activity Score (NAS) and hepatic hydroxyproline content were elevated at 3 and 14 mo on FFD. Picrosirius red staining of liver sections revealed a filigree pattern of fibrillar collagen deposition with no cirrhosis at 14 mo on FFD. Nevertheless, 46% of animals bore one or more tumors on their livers confirmed as HCC in hematoxylin-eosin-stained liver sections. Receiver operating characteristic (ROC) curves analysis for serum levels of the HCC biomarkers osteopontin (OPN), alpha-fetoprotein (AFP) and Dickkopf-1 (DKK1) returned concordance-statistic/area under ROC curve of $\geq$ 0.89. These data suggest that serum levels of OPN (threshold, 218 ng/mL; sensitivity, 82%; specificity, 86%), AFP (136 ng/mL; 91%; 97%) and DKK1 (2.4 ng/mL; 82%; 81%) are diagnostic for HCC in a clinically relevant model of NASH.

**Introduction**

Given the diabetes, obesity and metabolic syndrome epidemics, non-alcoholic fatty liver disease (NAFLD) has reached epidemic proportions.[1,2] Left untreated, NAFLD, which starts as simple steatosis, can progress (Figure 1) to non-alcoholic steatohepatitis (NASH), NASH with increasing levels of fibrosis, cirrhosis and hepatocellular carcinoma, (HCC).[1-3] A number of studies has reported that cirrhosis, secondary to alcoholism or hepatitis C virus (HCV), is a major risk factor for HCC and liver-related death.[4-6] In fact, the American Association of Liver Disease (AASLD) recommends screening of cirrhotics every 6 mo for HCC.[7,8] Nevertheless, mounting evidence[9-12] suggests that NASH can also progress to HCC in the absence of cirrhosis (Figure 1). Most alarming, data from a number[9-13] of clinical studies indicates that ~50% of NASH patients presents with HCC without cirrhosis.

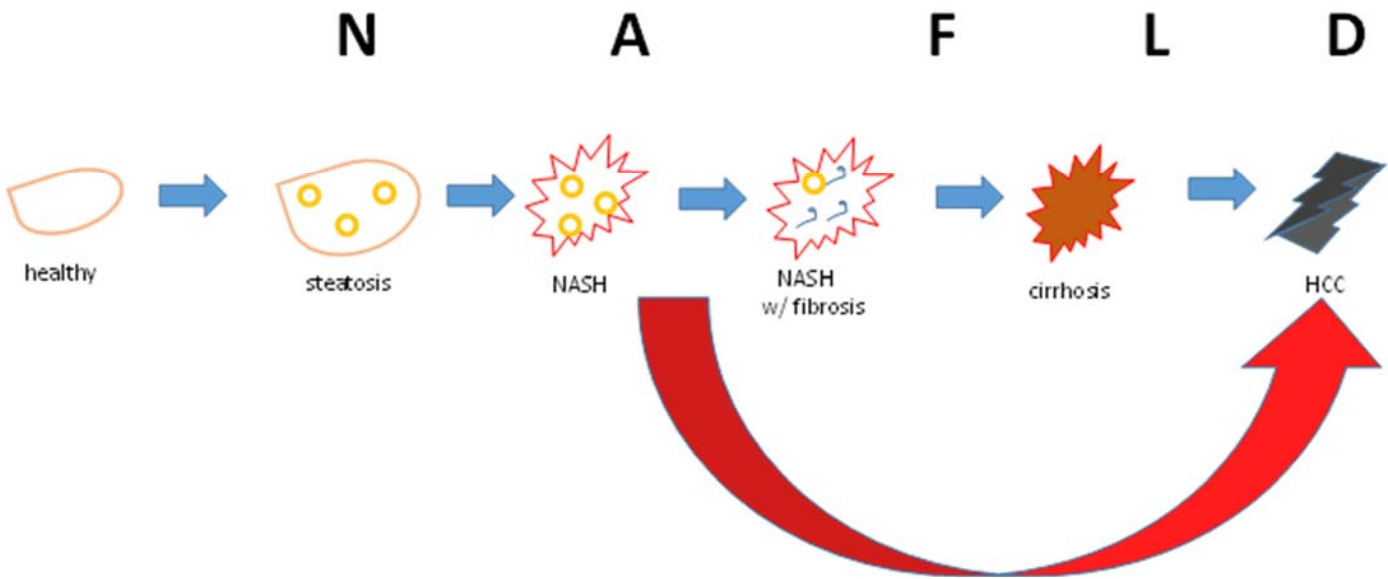

**Figure 1. NASH and HCC:** The spectrum of NAFLD is shown with classical progression (blue block arrows) from simple steatosis to NASH to NASH with fibrosis, cirrhosis and HCC. Nevertheless, it is becoming evident that NASH can lead to HCC (red arrow) in the absence of cirrhosis, and in certain cases, prior to the development of even advanced fibrosis.

Liver biopsy-microscopic evaluation remains the gold standard for diagnosis of HCC[9]. Given that seeding of the tumor in the needle tract occurs only in 1–3% of biopsies[14] and taking into account the logistics, risks and patient compliance issues associated with liver biopsies, ultrasonography with or without serum alpha-fetoprotein (AFP), an HCC biomarker, is being used to diagnose HCC.[7-8] However, a major challenge associated with HCC surveillance in patients with NASH is that obesity compromises the completeness of an ultrasound examination of the liver.[13] This necessitates computed tomography or magnetic resonance imaging scans which are expensive, both in terms of costs and patient compliance. Results from liver imaging can also be confounded by the presence of benign incidentelomas such as hamartomas, resulting in incorrect diagnosis. Use of a single HCC biomarker such as AFP is also fraught with risk of misdiagnosis - in many instances HCCs do not produce AFP; on the other hand, this biomarker can be elevated in the absence of HCC. Patients with NASH-HCC often present with larger tumors and have a worse prognosis than those with HCC secondary to other etiologies[13]. Additionally, it is fully recognized[7,8,15] that therapeutic strategy and prognosis in HCC is linked to disease stage, with Barcelona Clinic Liver Cancer (BCLC) stage 0-A patients having better outcomes. Taken together, these findings suggest that HCC surveillance is urgently needed in the ~ 16 million NASH patients and the ~ 5 million NASH with fibrosis patients within the United States alone[16]. Furthermore, from the patient's and loved ones perspective, correct diagnosis of HCC is as important as early diagnosis.

In the present study, we employed a clinically relevant, mammalian model[17,18] of diet-induced liver disease to develop a framework for diagnosis of HCC in NASH without cirrhosis. Mice on standard lab chow or a fast food diet (FFD; a.k.a. Western diet) were followed for up to 14 mo. With each animal acting as its own control, serum level of a panel of HCC-related biomarkers was evaluated and compared with the absence or presence of HCC. Receiver operating characteristic (ROC) curves for serum levels of HCC biomarkers were generated and for biomarkers with concordance-statistic or area under ROC curve (AUROC) > 0.8, thresholds identified. A Boolean logic gate-linked diagnostic framework for HCC in his model of NASH is presented.

**Methods**

The animal protocol was reviewed and approved by the local institutional animal care and use committee. Mice had access to food and water *ad libitum* throughout the course of the in-life protocol.

**Fast Food Diet (FFD) Model of Liver Disease:** Adult male C57BL/6 mice (20-22 g) were randomized to standard lab chow (control) or FFD (rodent diet with 40 kcal% fat (mostly Primex), 20 kcal% fructose and 2% cholesterol; D09100301, Research Diets, NJ) for up to 14 mo. Animals were sacrificed at 3 or 14 mo into their diets. Each individually tracked animal acted as its own serum and liver control.

**NASH and HCC:** At sacrifice, livers were examined (gross) for the absence or presence of tumors by two independent observers. Several biopsies were obtained from each liver and fixed in formalin (10%); whenever tumors were present, both tumors and remote liver were biopsied. Hematoxylin-eosin (H&E)-stained liver sections were studied under a microscope by blinded trained observers for evaluation of both, NAFLD Activity Score (NAS; worsening 0-8 scale; 0-3 steatosis; 0-3 inflammation; 0-2 hepatocyte ballooning)[19] and HCC.[20,21] NAS was averaged from three independent observers whereas HCC diagnosis required confirmation from both gross liver and microscopic observation by 2 trained and independent observers.

**Liver Fibrosis:** At sacrifice, total liver hydroxyproline, a surrogate for collagen, was quantified from liver biopsies using a colorimetric assay as described previously[22]. Picrosirius red staining of liver sections followed by semiquantitation (Bioquant) of extracellular fibrillar collagen[22] was performed by a blinded observer.

**Serum-based Biomarkers:** Commercially available enzyme-linked immunosorbent assay (ELISA) kits for murine serum samples were used for determination of biomarker levels. In many instances, following a pilot study, serum samples had to be diluted (saline) and the ELISAs rerun so that sample levels remained within the standard curve. Table 1 lists the biomarkers evaluated and information on the ELISA kit. Aspartate transaminase (AST) and alanine transaminase (ALT) were measured (Northwell Health, NY) in serum samples from the 3 mo cohort.

**Table 1. Biomarker Query Set:** Based on literature[23-28], levels of 7 biomarkers circulating in blood were evaluated.

| Biomarkers | ELISA Kit |
|---|---|
| alpha fetoprotein (AFP) | MAFP00; R&D |
| lens culinaris agglutinin A-reactive fraction of AFP (AFP-L3) | MBS724605; MYBioSource |
| des γ carboxyprothrombin (DCP aka PIVKA-II) | MBS2516006; MYBioSource |
| glypican -3 | MBS705612; MYBioSource |
| osteopontin | MOST00; R&D |
| golgi protein-73 (GP73) | MBS024709; MYBioSource |
| Dickkopf-related protein 1 (DKK1) | DY1765; R&D |

**Data Analysis**: Data are presented as mean ± standard error or mean. Student's T-test or one-way analysis of variance followed by Tukey's post-hoc test was used to compare data between groups. A $p < 0.05$ was deemed significant. ROC curves were generated in Excel using Pivot Tables and AUROC, sensitivity ($S_n$), specificity ($S_p$) and cutoffs/thresholds calculated.[29]

# Results

## Gross Observations

Of the 24 animals randomized to FFD for 14 mo, 11 animals, i.e. 46%, exhibited one or more tumors (Figure 2) on their livers with subsequent confirmation of HCC under microscope.

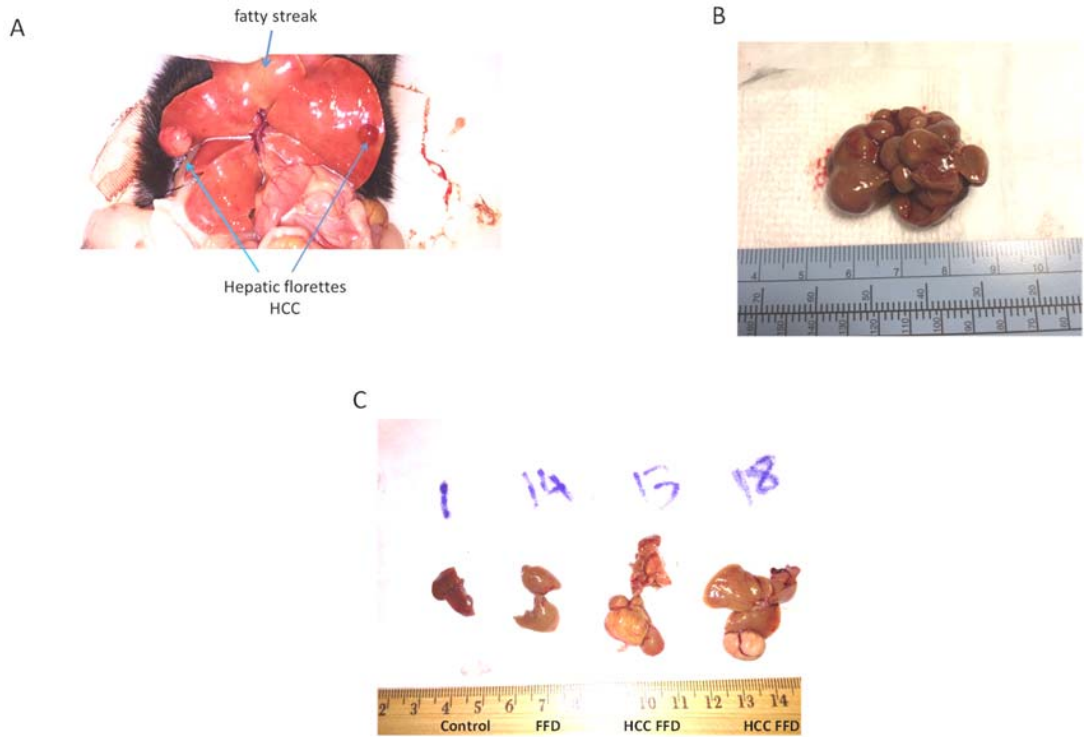

**Figure 2. Hepatic Tumors with FFD:** (A, B) Many animals on an FFD for 14 mo exhibited one or more tumors on the liver. (C) Representative livers from 14 mo - control (#1) and FFD (#s 14, 13 and 18) cohorts are shown. Large tumors are visible on the livers of #s 13 and 18.

Table 2 stratifies the animals entered into the study on the basis of diet type, duration on diet and presence or absence of HCC. By 3 mo on FFD while livers had a distinct fatty phenotype there were no visible tumors. By 14 mo on FFD, approximately half the animals had visible liver tumors.

| n | Diet | Duration on Diet | HCC |
|---|------|------------------|-----|
| 8 | control | 3 mo | no |
| 32 | FFD | 3 mo | no |
| 12 | control | 14 mo | no |
| 13 | FFD | 14 mo | no |
| 11 | FFD | 14 mo | yes |

**Table 2. Liver Phenotype:** Of the 76 animals entered into the study, 11 of 24 animals from the 14 mo FFD cohort exhibited HCC positive liver tumors.

At both 3 mo and 14 mo on FFD, animals presented with increased liver mass and liver:body mass ratio with these values being highest in livers presenting with tumors (Figure 3).

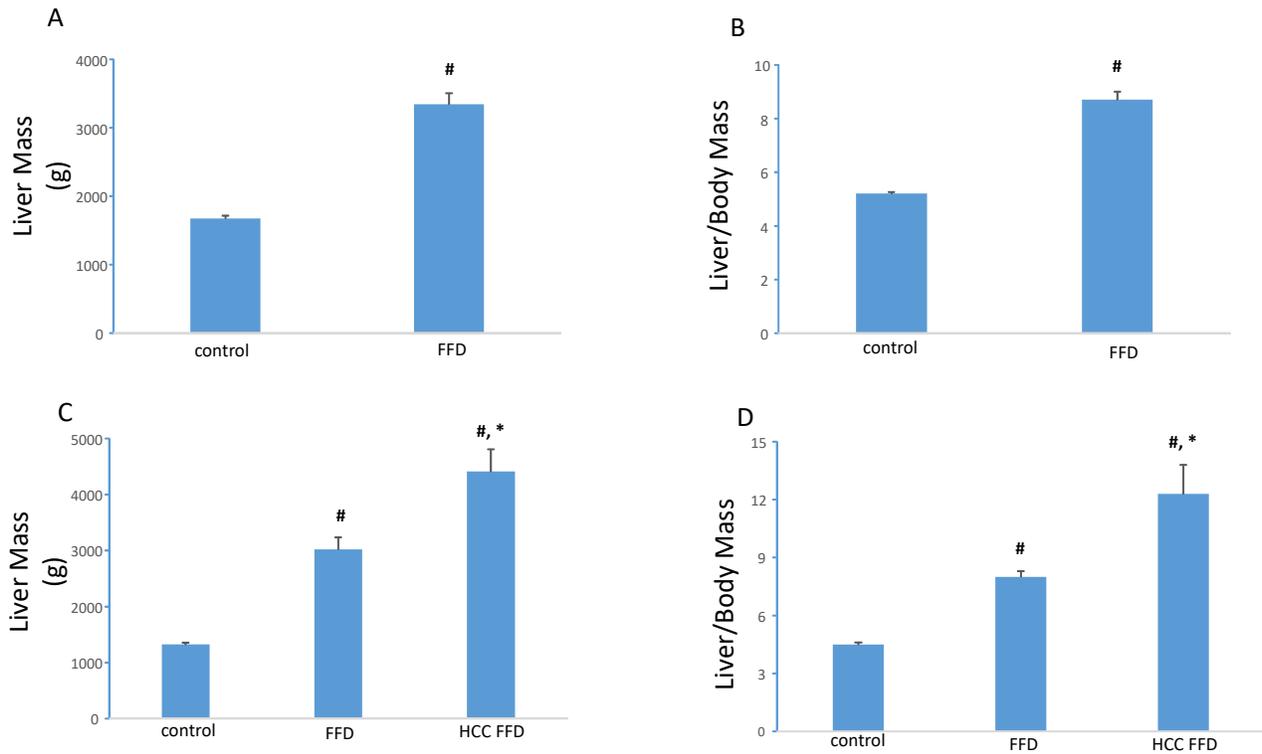

**Figure 3. Liver Phenotype:** Animals randomized to FFD exhibited increased hepatic mass and liver to body mass ratio. This was evident by 3 mo (A, B) and at 14 mo (C, D). 14 mo-FFD animals with liver tumors (HCC FFD) had the highest values for these variables. #, $p < 0.01$ vs. control; * $p<0.01$ vs. FFD

Analysis of ROC curves for liver mass and liver to body mass ratio yielded AUROCs of 0.83 and 0.82, respectively (Figure 4). Clinically, these endpoints are of little or no diagnostic value and thresholds were not computed.

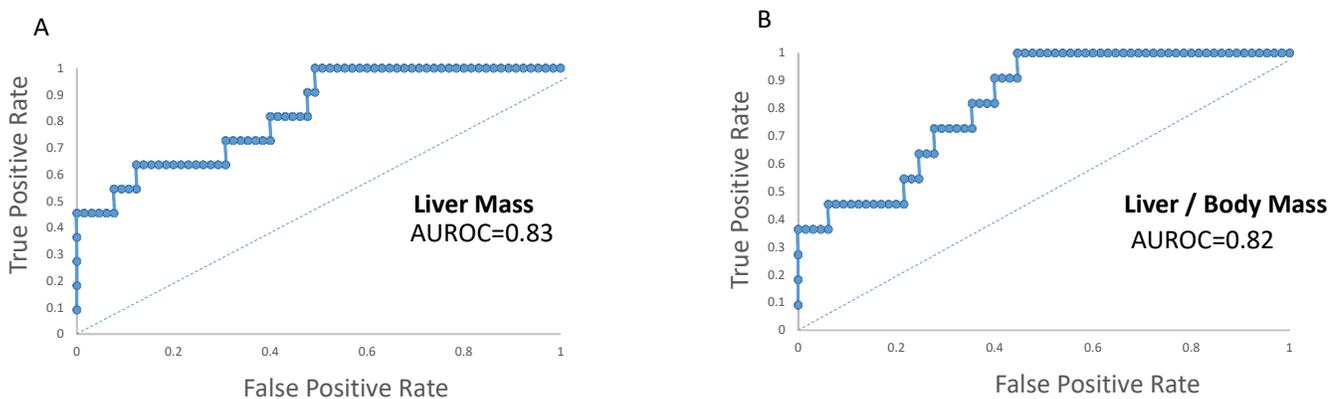

**Figure 4. Liver Phenotype and HCC:** AUROCs of liver mass (A) and liver to body mass ratio (B) to diagnose HCC were 0.83 and 0.82, respectively.

**Liver Histopathology and Liver Function Tests.**

Microscopic observation (Figure 5) of H&E stained liver sections from the FFD cohort showed hallmark characteristics of NASH. In fact, compared to the control diet cohort, NAS (Figure 5) was elevated at both 3 mo and 14 mo in the FFD cohort. Within the 14 mo FFD cohort, livers that presented with tumors on gross observation exhibited mixed pattern trabecular and acinar HCC (Figure 5). Comparing the 14 mo diet cohorts, the HCC livers had the highest NAS (Figure 5). Determination of liver function tests at the 3 mo time period demonstrated elevated AST (142±6 vs.69±17; p<0.01) and ALT (110±15 vs. 47±9; p<0.05) in the FFD vs. control diet cohort.

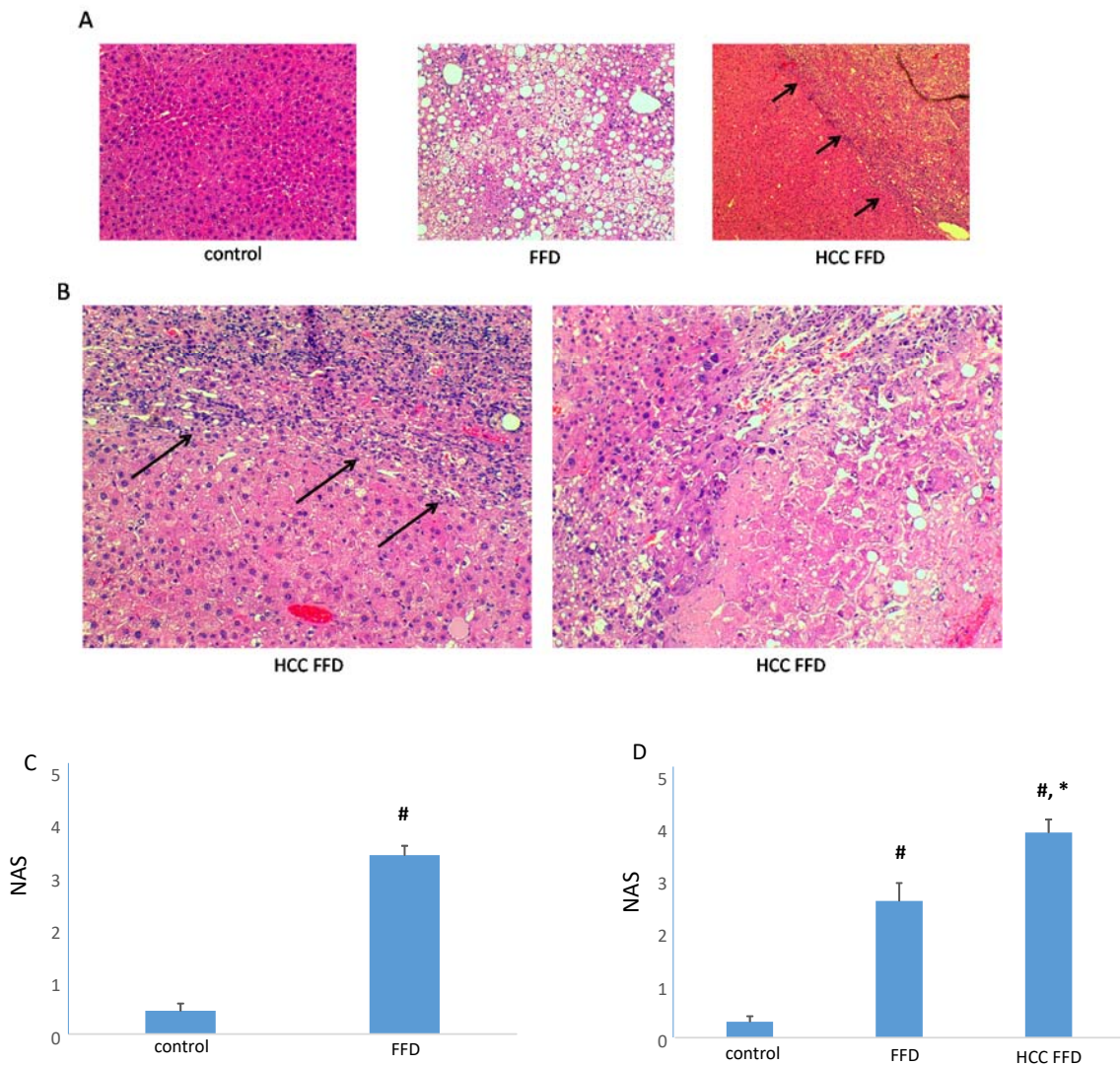

**Figure 5. FFD Model of Liver Disease:** (A) H&E-stained sections (4X) from representative 14 mo control diet or 14-mo FFD livers. Livers from the FFD cohort exhibited characteristic features of NASH. Some of the FFD livers showed HCC characterized by a demarcated zone (arrows) across which were multiple proliferating nucleii. (B) Representative livers (20X) from animals on FFD for 14 mo showing multiple proliferating nuclei (arrows), characteristic of HCC. These HCC livers also exhibited steatosis and inflammation. (C) Increased NAS was evident in livers of animals on FFD for 3 mo. (D) At 14 mo, livers presenting with HCC had the highest NAS. #, p< 0.01 vs. control; * p<0.01 vs. FFD

Next, we determined this model of NASH is accompanied by increasing levels of liver fibrosis. Animals randomized to FFD for 3 mo exhibited a 50% higher liver hydroxyproline content, a marker of fibrosis, compared with animals randomized to the control diet (Figure 6). At 14 mo, animals on FFD, but free of HCC, had >4-fold higher liver hydroxyproline content than animals on a control diet (Figure 6). Liver hydroxyproline content was highest in the 14 mo HCC FFD cohort, 10-fold greater than the 14 mo control diet cohort. The AUROC for liver hydroxyproline to diagnose HCC was 0.9 (Figure 6).

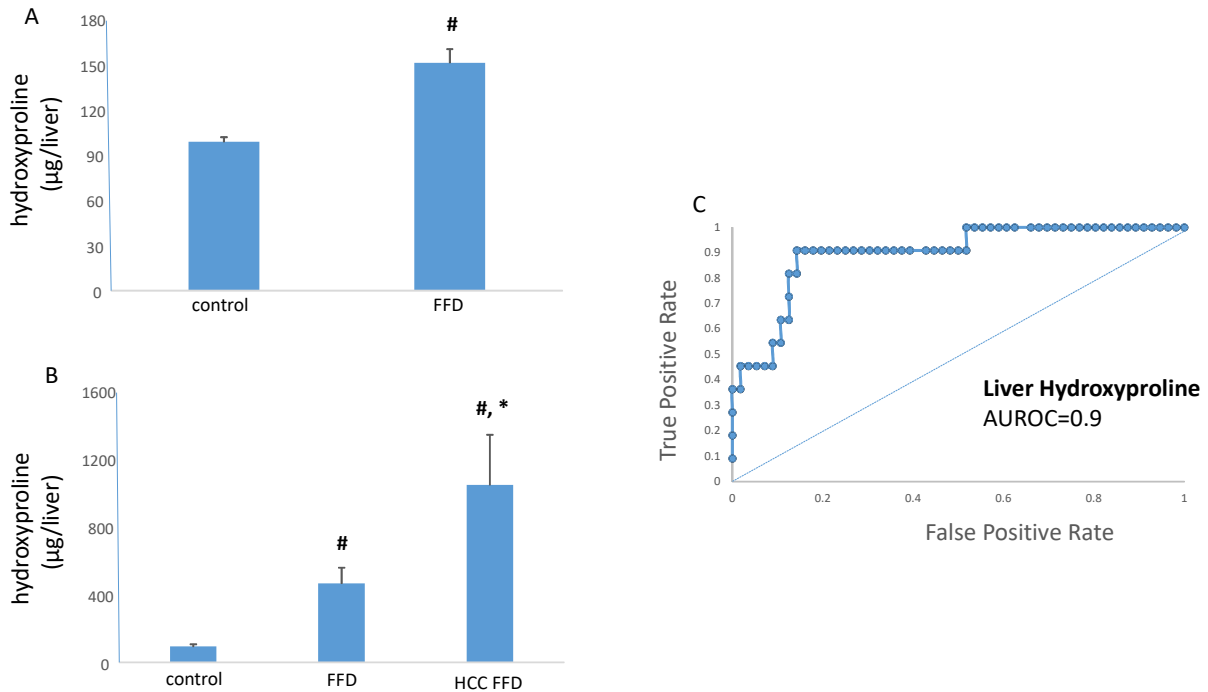

**Figure 6. NASH with Fibrosis and HCC.** Livers from animals on a control diet or FFD were examined for fibrosis using the marker hydroxyproline. A) 3 mo on FFD was associated with a 50% increase in liver hydroxyproline. B) By 14 mo, the HCC-free FFD cohort exhibited > 4-0 fold higher liver hydroxyproline content compared to the control group. At this timepoint, the HCC FFD cohort had the highest level of liver hydroxyproline. #, $p< 0.01$ vs. control; * $p<0.01$ vs. FFD. C) The AUROC of liver hydroxyproline as a diagnostic for HCC was 0.9.

Cirrhosis develops from fibrosis[30] although the two are pathologically distinct. At the fibrosis stage, the amount of collagen increases and the ratio of fibroconnective tissue versus liver cellular tissue increases; liver lobular structures are intact. Cirrhosis consists of two pathological features fibroconnective tissue hypertrophy and pseudolobule formation[30]. With cirrhosis, the liver's fundamental structure is deformed, and the framework of the liver begins collapse. To determine whether the HCC FFD cohort had progressed beyond fibrosis to cirrhosis, Picrosirius red-stained sections of livers were examined. While livers from the control diet cohort exhibited little, or no Picrosirius red staining, livers from the FFD cohort classified as HCC exhibited a filigree pattern of Picrosirius red staining (Figure 7), with a collagen proportionate area consistent with fibrosis but not cirrhosis[31].

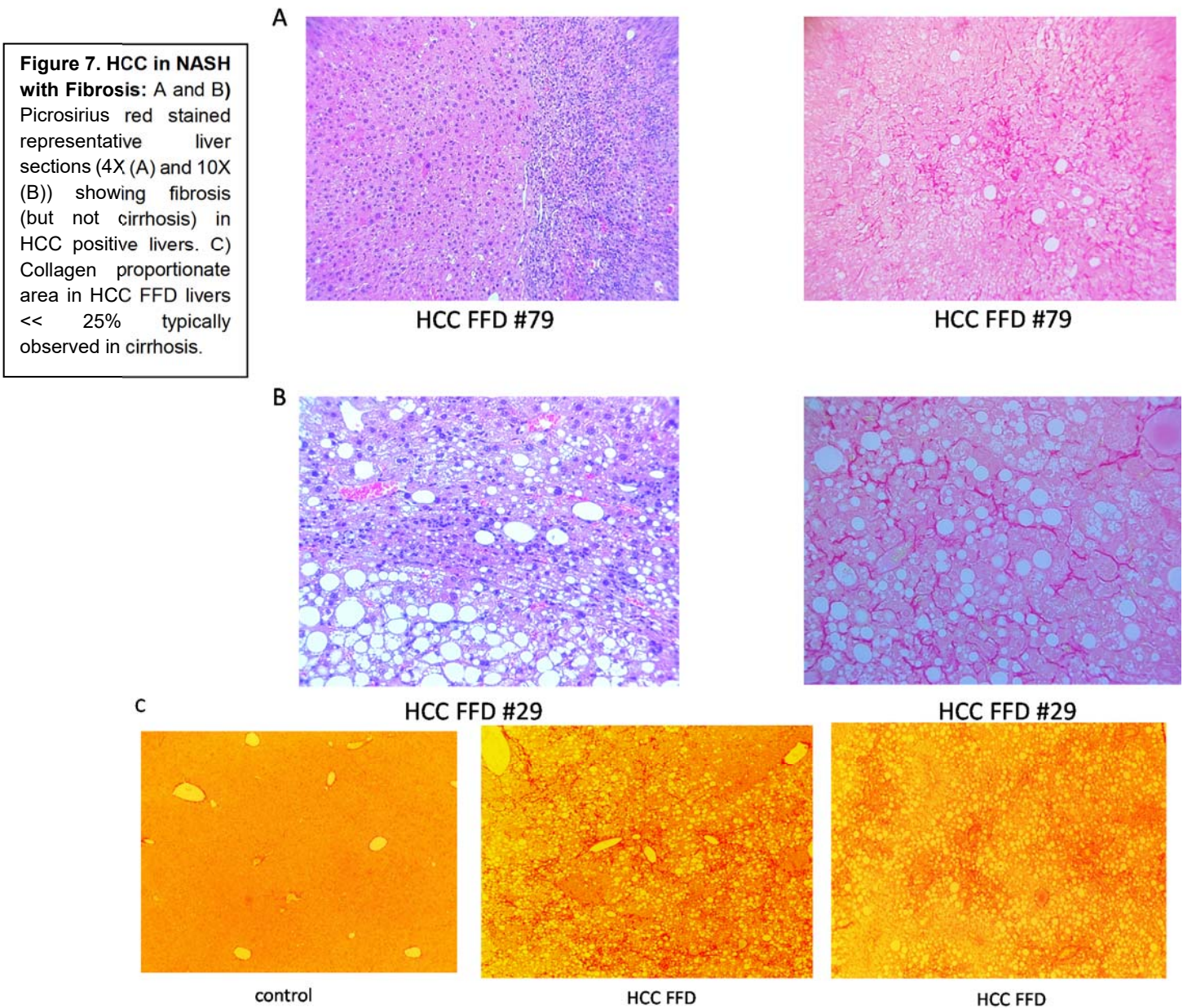

**Figure 7. HCC in NASH with Fibrosis:** A and B) Picrosirius red stained representative liver sections (4X (A) and 10X (B)) showing fibrosis (but not cirrhosis) in HCC positive livers. C) Collagen proportionate area in HCC FFD livers << 25% typically observed in cirrhosis.

## Serum HCC Biomarkers

Next, serum levels of biomarkers listed in Table 1 were measured in mice on control diets or FFD for 3 mo or 14 mo. With each animal acting as its own serum and liver control, ROC curves for the biomarkers were computed (Figure 8). In preliminary studies using a large and representative subset of serum samples, AFP-L3 levels << the standard curve and this biomarker was discarded. Of the biomarkers queried, osteopontin (OPN), alpha-fetoprotein (AFP) and Dickkopf-1 (DKK1) had AUROCs > 0.8 for HCC and $S_n$, $S_p$ and thresholds for these biomarkers were computed (Table 3).

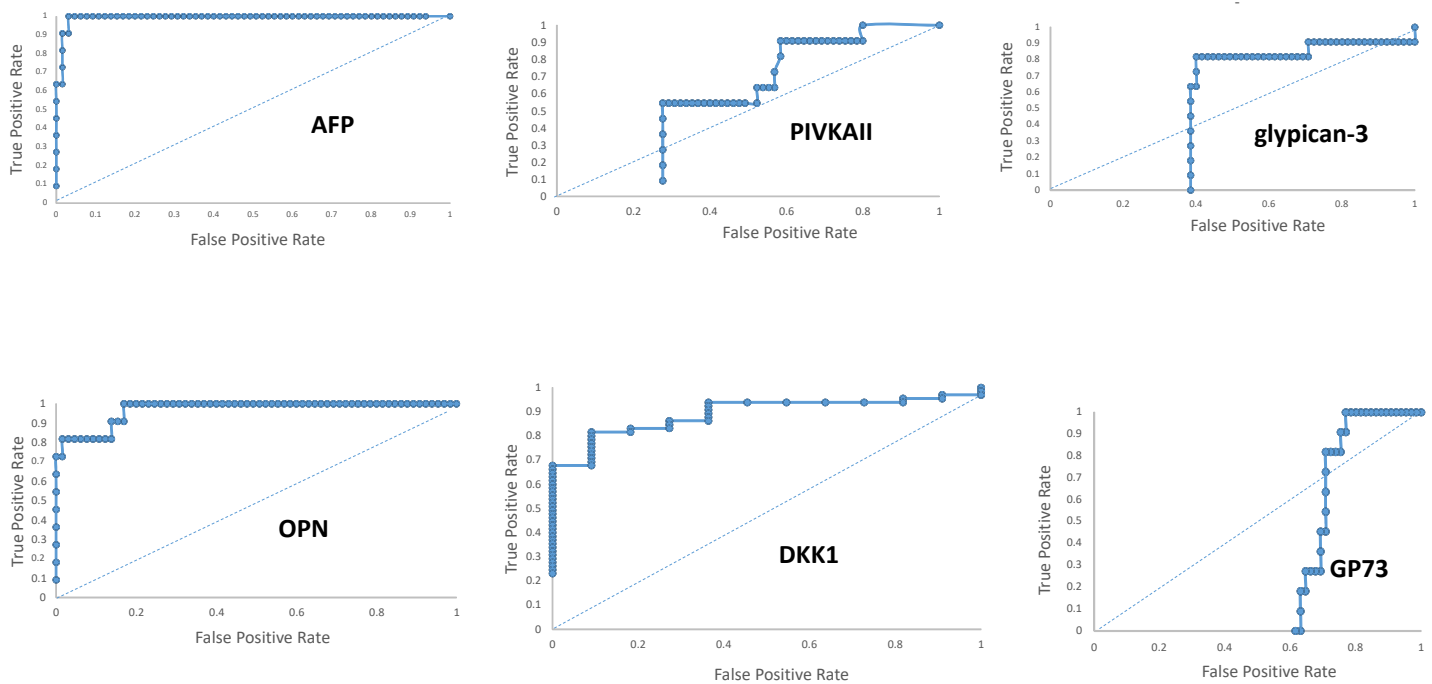

**Figure 11. ROC Curves for HCC Biomarkers**: ROC curves are shown for AFP, DCP/PIVKA II, glypican-3, OPN, DKK1 and GP73.

| Biomarker | AUROC | Sn (%) | Sp (%) | Threshold (ng/mL) |
|---|---|---|---|---|
| OPN | 0.97 | 82 | 86 | 218 |
| AFP | 0.99 | 91 | 97 | 136 |
| DKK1 | 0.89 | 82 | 81 | 2.4 |

**Table 3. Diagnostic Profile for OPN, AFP and DKK1.**

Measurement of serum levels of OPN, AFP and DKK1 provides a framework for diagnosing HCC in this diet-induced model of NASH (Figure 9).

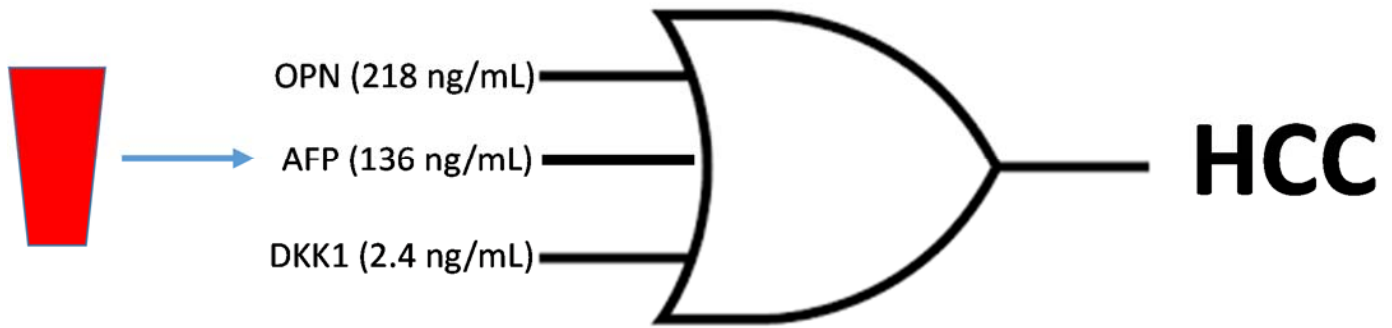

**Figure 8. Diagnostic Framework in NASH.** Serum levels of OPN or AFP or DKK1 can be used to diagnose HCC in a diet-induced model of NASH.

Discussion

In a murine model of FFD-induced NAFLD, animals developed NASH with increasing levels of fibrosis but not cirrhosis. Furthermore, approximately half the animals on FFD for 14 mo developed HCC. Analysis of a panel of serum-based biomarkers indicates that OPN, AFP and DKK1 are good diagnostics for HCC in NASH with AUCOCs of 0.97, 0.99 and 0.89, respectively and thresholds of 136 ng/mL, 218 ng/mL and 2.4 ng/mL, respectively.

HCC is the fifth most common tumor worldwide and the second most common cause of cancer-related death[7,8]. Its incidence is increasing rapidly, primarily due to the HCV epidemic in the United States, and the hepatitis B virus epidemic in the Far East[7,8,26]. Fatty liver disease can also lead to HCC. With ~ 30% or adults in the United States having some steatosis, and 3-12% of adults in the United States having NASH, it is estimated that fatty liver disease-related HCC will overtake all other causes of primary liver cancer with the next 2 or 3 decades[16].

Within the continuum of liver disease, cirrhosis carries increased risk for HCC. In fact, the American Association of Liver Disease recommends surveillance of adults with cirrhosis every 6 months because early detection improves overall survival[7,8]. Very worrisome has been the recent increase of HCC in patients with NASH but no cirrhosis[9]. In a review[10] of the tumor board database at Brooke Army Medical Center (BAMC) 13% of the HCC cases occurred in non-cirrhotic NAFLD/NASH. In a study[32] by Dyson and colleagues characterizing the demographics of HCC from 2000 to 2010 in the region surrounding Newcastle-upon-Tyne in North East England, ~30% of the HCCs diagnosed occurred in the absence of cirrhosis, findings echoed by Paradis et al[11] who reported development of HCC in patients with metabolic syndrome and the absence of significant liver fibrosis. In a clinical study[12] by Ertle and colleagues, ~ 42% of individuals with NAFLD-NASH-HCC had no evidence of cirrhosis. Following a comprehensive survey of the clinical literature, Kolly and Dufour reported[13] a 50% prevalence of HCC in the context of NASH without cirrhosis. These percentages translate to staggering numbers given that ~16 million people in the United States alone have NASH and ~5 million of those have NASH with advanced fibrosis[16]. Prognosis with HCC is often dismal with BCLC Stages 0 and A associated with better outcomes and BCLC Stages B-D deemed non-curative[7,8,15]. Patients with NASH-related HCC often present with large tumors and have a worse prognosis than those with other etiologies[13]. A retrospective analysis of the

Surveillance, Epidemiology and End Results (SEER)-Medicare database, 50% of viral-hepatitis-infected patients with HCC died after one year, whereas 61% of the NAFLD patients with HCC died after one year[33]. Needless to say, there is an emergent need for HCC surveillance in the NASH patient.

In the present study, we tracked animals randomized to a control or FFD for up to 14 mo in an attempt to simulate, at least, in part, fatty liver disease. The FFD included fat, cholesterol and fructose, a recipe that is associated with steatohepatitis[17,18]. Indeed, within 3 mo on this diet, animals had developed hallmark characteristics of NASH including elevated liver function tests. Some liver fibrosis, evidenced by a 50% increase in liver hydroxyproline content, was evident at 3 mo on this diet whereas by 14 mo on this diet animals presented with more advanced liver fibrosis. Picrosirius red is a marker of fibrillar collagen and use of this stain demonstrated a filigree network of collagen deposition, consistent with the development of liver fibrosis in this model. A salient and clinically relevant feature of this model is progression to HCC. While HCC has been reported[34,35] in other models of diet-induced liver disease, those models typically combine a chemical such as streptocozocin or diethyl nitrosamine as a means to accelerate and aggravate liver pathology, often resulting in ~100% incidence of HCC in 12- 20 weeks. The FFD model described herein is designed to mimic a Western diet and has a more natural progression to HCC, similar to that observed clinically. The other salient and clinically relevant aspect of this model is the appearance of HCC in the absence of cirrhosis. In fact, at 14 mo on FFD, while animals developed liver fibrosis none of the animals had progressed to a cirrhotic phenotype. Absence of cirrhosis was evidenced by the filligree pattern of Picrosirius red staining, and is a finding consistent with other reports[17,18] using fatty diet models. In the present study, of the animals on FFD for 14 mo, 46% developed HCC, a percentage in excellent agreement with some of the clinical reports referenced above. ROC curve analysis showed that HCC is present in animals with increased liver mass, increased liver to body mass ratio and increased liver fibrosis, in as much as liver hydroxproline content is a marker of the same. These data suggest that in this murine FFD model, NASH with fibrosis results in HCC and supports the notion that this model is suitable for mimicking the clinical findings of NASH-related HCC in the absence of cirrhosis.

Utilization of serum biomarkers has played a major role not only surveillance strategies in high-risk populations and achieving a diagnosis of HCC, but also in risk stratification and prediction of recurrence following initial therapy[23-28]. AASLD guidelines recommend screening for HCC using ultrasound with or without serum AFP. Several serum markers for HCC have been identified in addition to AFP for diagnosis of HCC. These biomarkers include a fucosylated isoform of AFP reactive to *Lens culinaris* agglutinin, known as AFP-L3, glypican-3, OPN, GP73, the abnormal prothrombin vitamin K absence-II (PIVKA-II) also known as Des-gamma carboxyprothrombin (DCP) and DKK1, among others[23-28]. Many of these biomarkers are in clinical testing but other than AFP none has clinical approval. Furthermore, each of these biomarkers, including AFP, suffers from suboptimal $S_n$ and $S_p$ making its use as a standalone diagnostic unreliable. HCCs are not homogenous; each is different. Thus, some HCCs may have normal or only mildly elevated levels of AFP, but high levels of AFP-L3. Similarly, some of the tumors may have high levels of PIVKAII/DCP while having normal levels of AFP. Levels of many of these biomarkers including AFP are elevated in the setting of fatty liver and liver regeneration which can lead to a false positive diagnosis for HCC. With current data suggesting that no single biomarker alone is likely to have optimal $S_n$ and $S_p$ for the detection of HCC, particularly during early disease, combining biomarkers might represent a mechanism to improve diagnosis. In fact, gastroenterologists in Japan and the United Kingdom rely on BALAD-2 and GALAD calculators[27] for diagnosing HCC albeit the scores are a function of disease etiology and are influence by regional differences.

Following a comprehensive survey of the literature[23-28], we queried a panel of serum-based HCC biomarkers (Table 1) to develop a diagnostic framework for HCC in our clinically relevant model NASH. A sample set of 76 animals on 2 different diets and 2 different timepoints was used in this study with each animal acting as its own control. HCC was diagnosed by the presence of tumors on the liver coupled with histopathological confirmation of disease. Based on ROC analysis, 3 biomarkers, OPN, AFP and DKK1, and their thresholds, were identified with excellent diagnostic potential for HCC in NASH. Our empirical findings led us to a quantitative framework for diagnosis of NASH-related HCC without cirrhosis, which, to the best of our knowledge, is the first such description. The other biomarkers, viz. AFP-L3, PIVKAII/DCP, glypican-3 and GP73 performed poorly as diagnostics in this diet-induced model of NASH-related HCC; including the GALAD subset, AFP-L3 and PIVKAII/DCP. The reason(s) for their poor performance is beyond the scope of this study albeit it has been argued[36] that a biomarker that consistently misdiagnoses disease can still find use by reversing (1-AUROC) the diagnosis.

These findings are clinically relevant. Most importantly, risk for misdiagnosis of HCC is distributed amongst 3 potentially independent diagnostics. Use of serum samples enables repeated sampling with minimal discomfort to the subject while minimizing the need for imaging studies. In addition to their use as a diagnostic, these biomarkers can be used to determine the efficacy of treatment strategies. Nevertheless, there are some limitations to these findings. It remains to be determined whether these findings are applicable to other models of NASH and to species other than the mouse. Second, tumor burden has not been correlated with biomarker levels. It would be important to conduct a longitudinal study coupling serum biomarker analysis with high resolution liver imaging, to determine how early HCC can be diagnosed in the setting of NASH and without the need to sacrifice the animals. Finally, the biomarkers queried is merely a subset of the HCC-related biomarkers described in the literature[23-28] opening the possibility of expanding our findings. Nevertheless, this study represents, in our opinion, the initial step in the clinical diagnosis of NASH-HCC in the absence of cirrhosis.